\documentclass{iopart}

\usepackage{amssymb}
\usepackage{amsthm}

 \newtheorem{Th}{Theorem}

\def\be{\begin{equation}}
\def\ee{\end{equation}}
\def\bea{\begin{eqnarray}}
\def\eea{\end{eqnarray}}

\def\p{\partial}

\def\l{\lambda}

\def\vfi{\varphi}

\def\wt{\widetilde}

\def\pt{{$\cal PT$}}
\def\ptt{{$\cal PT~$}}
\def\c{{$\cal C~$}}
\def\l2{{$L^2(-\pi ,\pi )$ }}
\def\cpt{{$\cal CPT$}}

\begin{document}

 \large

\title {Is the \cpt-norm always positive?}

\author{
 Boris F Samsonov$^{\dag}$ and
 Pinaki Roy$^{\S}$
}

\address{\dag\ Physics Department of Tomsk State
 University, 634050 Tomsk, Russia}

\address{\S \, Physics
\symbol{38} Applied Mathematics Unit
Indian Statistical Institute
Kolkata - 700 108}

\ead{\mailto{samsonov@phys.tsu.ru},
\mailto{pinaki@isical.ac.in}
 }

\begin{abstract}
\baselineskip=16pt
\noindent
We give an explicit example of an exactly solvable \pt-symmetric
Hamiltonian with the unbroken \pt\ symmetry which has one eigenfunction
with the zero \pt-norm.
The set of its eigenfunctions is not complete in corresponding
Hilbert space and it is {\it non-diagonalizable}.
In the case of a regular Sturm-Liouville problem
any {\it diagonalizable} \pt-symmetric Hamiltonian
with the unbroken \pt\ symmetry has a complete set of
 positive \cpt-normalazable eigenfunctions. For non-diagonalizable
 Hamiltonians a complete set of \cpt-normalazable functions is
 possible but the functions belonging to the root subspace
 corresponding to multiple zeros of the characteristic determinant
 are not eigenfunctions of the Hamiltonian anymore.
\end{abstract}


\medskip
\medskip








{\bf 1.}
In recent years it has been shown that non Hermitian Hamiltonians,
in particular the \pt-symmetric ones may have real eigenvalues \cite{Bender}.
 This has given rise to the possibility of constructing
 a complex extension of quantum mechanics \cite{BBJ}.
Before the
discovery of the \c operator \cite{BBJ} the main difficulty in constructing a
self-consistent complex extension of quantum mechanics was
the presence of negative \pt-norms for some \pt-symmetric
Hamiltonians.
Using the \cpt operation a new norm was defined \cite{BBJ}
and it was shown to be positive for some models.
In this Letter we would like to stress that this is true only if
the \pt-norm is non-zero. Here we shall consider an explicit example of an
exactly solvable \pt-symmetric Hamiltonian one of whose  eigenfunction
has the zero \pt-norm thus proving that such a situation may
occur in a \pt-symmetric Hamiltonian. Further, using the fact that
the \ptt operation applied to an eigenfunction of the given
Hamiltonian may only change its sign
resulting in the property that the \cpt\ operation inverts the sign
of a \ptt norm if it is negative, we conclude that application of the
\cpt\ operation
to an eigenfunction of a \ptt symmetric Hamiltonian
is equivalent to first going to an eigenfunction of the
adjoint operator and then taking the complex conjugation provided
the given solution is normalized properly.
Subsequently we show that
for a regular Sturm-Liouville problem the
zero \cpt-norms may appear only if the characteristic determinant
has multiple zeros. In this case the system of eigenfunctions is not complete
 in
corresponding Hilbert space
implying that the Hamiltonian related with such
a problem is {\it non-diagonalizable}. We show that both for
diagonalizble and non-diagonalizable Hamiltonians one is able to
define a Hilbert space with a positive \cpt-norm but in the later
case the basis functions corresponding to a degenerate root
subspace are not eigenfunctions of the Hamiltonian.

{\bf 2.}
In this part of our note we present an example of an exactly solvable
 \pt-symmetric Hamiltonian having one eigenfunction with the zero
 \pt-norm.

Let us consider the following Sturm-Liouville problem in the interval
$[-\pi ,\pi ]$:
 \be\label{1}
(-\p_x^2+V(x)-E)\psi=0
\ee
with the zero boundary conditions
\be\label{bc}
\psi(-\pi)=\psi(\pi)=0\,.
\ee
We would like to consider the following \pt-symmetric potential
\be\label{V}
V(x)=-\frac{6}{(\cos x+2i\sin x)^2}\,.
\ee
It is not difficult to check that for a given
 $E=k^2\in \Bbb C$ Eq. (\ref{1}) with the potential (\ref{V})
  has the following
solutions
\be\label{psi12}
\psi_1=e^{ikx}[2i-ki+\frac{3}{2i+\cot x}]\qquad
\psi_2=e^{-ikx}[2i+ki+\frac{3}{2i+\cot x}\,]\,.
\ee
For all $k\ne 1$ they are linearly independent since their
Wronskian is $2ik(k^2-1)$. For $k=1$ one can choose
\be
\psi_1=\frac{1}{\cos x+2i\sin x} \qquad
\psi_2=\frac{5\sin(2x)-4i\cos(2x)-6x}{\cos x+2i\sin x}\,.
\ee
The Wronskian of these functions is equal to 4.

It is evident that the zero boundary conditions may be satisfied
if and only if  $\Delta=\Delta(E)=0$ where
\be\label{D}
\Delta=\left|
\begin{array}{cc}
a_1 & a_2 \\ b_1 & b_2
\end{array}
 \right|
 \qquad a_{1,2}=\psi_{1,2}(-\pi )
\qquad b_{1,2}=\psi_{1,2}(\pi )
\ee
is the characteristic determinant.
From here for $k\ne 1$ one yields
\be\label{a}
(k^2-4)\sin(2k\pi)=0\,.
\ee
This corresponds to the purely real spectrum $k=k_n=n/2$
\be
E_n=n^2/4 \qquad n=1,3,4,5,\ldots .
\ee
It may be pointed out that all roots of equation
(\ref{a})
are simple except for $E=k^2=4$ which is a double root
and it will be shown that this result has important
consequences. For $k=1$, $\Delta\ne 0$
meaning that $E=1$ ($n=2$) is not a
spectral point.
We would like to note that the whole spectrum is simple
(i.e. there is only one eigenfunction for every eigenvalue)
including the point $E=4$ and the existence of the double root of
the equation $\Delta(E)=0$ is not related with the (non-)degeneracy
of the eigenvalue $E=4$.
This can be easily seen from
(\ref{psi12}). For instance at $k=4$ $\psi_1(\pm\pi)=0$ but
$\psi_2(\pm\pi)\ne 0$.

The eigenfunctions are
\bea\nonumber
\psi_n=\frac{
[(16-n^2)\cos x-2i(n^2-4)\sin x]\sin[\frac n2(\pi+x)]
-6n\sin x\cos [\frac n2(\pi+x)]}%
{\cos x+2i\sin x}
\,.
\eea
They are ${\cal PT}$ orthogonal,
(we remind that \pt$\psi_n(x)=\psi_n^*(-x)$)
and  it is not difficult to find their ${\cal PT}$-norm, so
\be
\int_{-\pi}^\pi \psi_n(x)\,[{\cal PT}\psi_m(x)]dx=
{\pi}\,(-1)^{n+1}\, (n^2-4)\,(n^2-16)\delta_{nm}\qquad n\ne 2\,.
\ee
Thus we see that the \pt-norm of $\psi_4$ is null.
This means that if
one defined the \cpt-inner product by
  redefining the \pt-inner product
in a way that the vectors with a negative  \pt-norm would
become the vectors with a positive \cpt-norm
(in our case would be $\|\psi_n\|_{\cal CPT}^2=\pi (n^2-4)\,(n^2-16)$,
$n=1,4,5,\ldots$ and $\|\psi_3\|_{\cal CPT}^2=35\pi$),
the vectors with the zero \pt-norm
($\psi_4$ in our example)
 would still remain
as vectors with the zero \cpt-norm and the metric
of such a space would be neither negative nor positive.
Another interesting observation is
${\cal PT}\psi_n =(-1)^{n-1} \psi_n$, $n=1,3,4\ldots$,
but since the vector with $n=2$ is
missing the $\pm$ signs do not alternate
for two adjacent points of the spectrum $n=1$ and $n=3$.

An important property of a spectral problem such as
the one given in (\ref{1}-\ref{V})
is that the set of eigenfunctions $\{\psi_n\}$
is not complete in $L^2(-\pi,\pi)$.
Nevertheless, it is remarkable that one can find the missing functions
and enlarge the set of eigenfunctions till a set complete in
$L^2(-\pi,\pi)$
and these missing functions are related just with the multiple
roots of the equation $\Delta(E)=0$. We shall show now that in our
particular case the single missing function
 is
related not with the missing value of $n=2$
but with the eigenfunction
\be\label{psi4}
\psi_4=\frac{-24e^{2ix}\sin x}{\cos x+2i\sin x}
\ee
corresponding to the double
root of the equation $\Delta(E)=0$. To find this function we introduce a special
solution $\psi(x,k)$, $E=k^2$ of the spectral problem (\ref{1}-\ref{bc})
such that $\psi(-\pi,k)=0$ fixing the normalization by the
condition $\psi'(-\pi,k)=1$. (By the prime we denote the
derivative with respect to $x$.)
It can easily be found with the help of the
solutions (\ref{psi12})
 to see that
\be\label{psiseries}
\psi(\pi,k)=\frac{(k^2-4)\sin(2k\pi)}{k(k^2-1)}
\ee
and the equation $\psi(\pi,k)=0$ has exactly the same roots as
 $\Delta(E)=0$ (see equation (\ref{a})).
 In particular all roots are simple except for $k=2$ which is a
 double root.
(Notice that $\psi(x,\frac n2)$ may differ from $\psi_n$ only by a
constant factor.)
By this reason
its derivative with respect to $k$,
$\dot\psi(x,k)\equiv \p\psi(x,k)/\p k$,
at $k=2$
\be\label{AssFun}
\dot\psi(x,2)={\textstyle\frac 1{12}}
[12i\pi -7+12ix+8e^{-2ix}-e^{-4ix}]{\psi(x,2)}
\ee
 satisfies the zero boundary conditions,
$\dot\psi(\pm\pi,2)=0$ also.
It evidently is linearly independent
with the function (\ref{psi4}) and \pt-orthogonal with $\psi_n$,
$n=1,3,5,6,7\ldots$ which may be checked by the direct
calculation meaning that it is linearly independent with
 the set of eigenfunctions $\{\psi_n\}$.
It follows from equation (\ref{1}) that it satisfies
the inhomogeneous equation
\be\label{psidot}
[-\p_x^2+V(x)-4]\dot\psi(x,2)=4\psi(x,2)\,.
\ee
The function $\dot\psi(x,2)$
is called {\em associated function} with the eigenfunction $\psi(x,2)$
(see e.g. \cite{Naimark,Marchenko}). It can be proven
(see e.g. \cite{Marchenko}, theorem 1.3.1)
 that the set
$\{\psi_n\}$, $n=1,3,4,5\ldots$ supplemented with $\dot\psi(x,2)$
or equivalently with
\be\label{fi4}
\vfi_4=\frac{12ixe^{2ix}-e^{-2ix}+8}{\cos x+2i\sin x}\,\sin x\,.
\ee
is complete in $L^2(-\pi,\pi)$. One can notice that
$\psi_4$ and either $\vfi_4$ or $\dot\psi(x,2)$ form a basis in
the two-dimensional root subspace ${\cal L}_4$
corresponding to the energy $E=4$ and both they
satisfy the homogeneous equation
\be\label{eqfi4}
[-\p_x^2+V(x)-4]^2\psi(x)=0 \qquad
\psi(-\pi )=\psi(\pi)=0\,.
\ee
For $\psi_4$ this follows from (\ref{1}) and for $\dot\psi(x,2)$ one
should take into account equation (\ref{psidot}) also.

Let us introduce a short notation to the integral
\[
\int_{-\pi}^\pi\psi_n(x)\psi_m(x)dx\equiv (\psi_n,\psi_m)\,.
\]
Then by the direct calculation one can find that
 $(\vfi_4,\vfi_4)=-44\pi$ and $(\vfi_4,\psi_4)=-96\pi$.
Now in the root subspace ${\cal L}_4$ one can choose the basis
we denote $\xi_2$ and $\xi_3$ such that
$(\xi_n,\xi_m)=\delta_{nm}$, $n,m=2,3$,
\[
\xi_2=i\psi_4\sqrt{(\vfi_4,\vfi_4)}/(\vfi_4,\psi_4)-
i\vfi_4/\sqrt{(\vfi_4,\vfi_4)}
\qquad
\xi_3=\vfi_4/\sqrt{(\vfi_4,\vfi_4)}
\]
and renormalize all other basis functions
\[
\xi_n=\psi_n/\sqrt{\pi(n^2-4)(n^2-16)}\qquad
n=1,5,6,7,\ldots\qquad
\xi_4=\psi_3/\sqrt{-35\pi}\,.
\]
So, the new basis $\{\xi_n\}$ has the following properties:
\[
{\cal PT} \xi_n=(-1)^{n-1}\xi_n\qquad (\xi_n,\xi_m)=\delta_{nm}
\]
\[
(-\p_x^2+V(x)-E_n)^2\xi_n=0\qquad \xi_n(\pm\pi)=0
\]
 with
$E_n=n^2/4$ for $n=1,5,6,7,\ldots$, $E_2=E_3=4$ and $E_4=9/4$
which readily follow from the properties of the functions
$\psi_n$ and $\vfi_4$.

Before ending this section we would like to point out that the
zero \pt-norm of $\psi_4$ is not accidental but is due to the
fact that the root $k=2$ of the equation $\psi(\pi,k)=0$ is double root.
Indeed, since $\psi(x,k)$ satisfies equation (\ref{1})
 one has
 \[
(k^2-{\widetilde {k}} {\,^2})\psi(x,k)\psi(x,\wt k)=
\frac{d}{dx}[\psi'(x,k)\psi(x,\wt k)-\psi(x,k)\psi'(x,\wt
k)]
 \]
 from which it follows that at $\wt k=n/2$
\be
\int_{-\pi}^\pi\psi(x,k)\psi(x,{\textstyle\frac n2})dx=
\frac{1}{{\textstyle\frac {n^2}{4}}-k^2}
\psi'(\pi,{\textstyle\frac n2})\psi(\pi,k)\,.
\ee
Noting that $\psi'(\pi,{\textstyle\frac n2})=(-1)^n\ne 0$
and taking into account (\ref{psiseries}) we conclude that for $k=2$
($n=4$), $(\psi_4,\psi_4)=0$,
and for $k\ne 2$, $(\psi_n,\psi_n)\ne 0$,
$n=1,3,5,6,7,\ldots$. This is a property which
does not depend on a particular choice of the potential $V(x)$ and
takes place for any
eigenfunction (if present) of the boundary value problem (\ref{1}-\ref{bc})
with the simple spectrum
corresponding to a double root of the equation
$\psi(\pi,k)=0$. In particular, any such an eigenfunction of a
\pt-symmetric Hamiltonian corresponding to a regular
Sturm-Liouville problem has the zero \pt-norm.

So, from this example we see that the set of eigenfunctions of the
problem (\ref{1}-\ref{V}) is not complete but may be completed.
In the next section we shall see
that such a situation though unacceptable from the
quantum mechanical viewpoint
is usual in the theory of ordinary linear differential operators
 and our example presents an elementary
illustration of known theorems \cite{Naimark,Marchenko}.
This is related to the fact that in the usual
quantum mechanics one always deals with {\em diagonalizable}
Hamiltonians while in complex
quantum mechanics this is not always so.

{\bf 3.}
Here we first recall some facts
from the theory of ordinary linear differential operators \cite{Naimark,Marchenko}
 and then show how a new ({\em dynamical}) inner product
in the space \l2 can be defined.
Everywhere we shall assume that the
spectrum of the boundary value problem of type (\ref{1}-\ref{bc}) is real as it
is in our example.

{\bf I.}
A boundary value problem similar to that given by (\ref{1}-\ref{bc})
with a complex-valued function $V(x)$
defines a non-selfadjoint operator $H$ in the Hilbert space \l2 with
the dense domain of definition consisting of all twice differentiable
functions vanishing at $x=\pm \,\pi$. The adjoint problem obtained
from (\ref{1}-{\ref{bc}) by replacing $V(x)$ with its complex
conjugate $V^*(x)$ defines the operator $H^+$ which is Hermitian adjoint to
$H$. Since $H$ has a real spectrum, $H^+$ has the same spectrum and
the known bi-orthogonality condition between the eigenfunctions of
$H$, $\psi_n(x)$, and those of $H^+$, $\wt\psi_{n'}(x)=\psi_{n'}^*(x)$,
 has the form
\be\label{ip}
\langle \wt\psi_{n'}|\psi_n\rangle \equiv
(\psi_{n'},\psi_n)=
\int_{-\pi}^\pi \psi_{n'}(x)\psi_n(x)dx=0 \qquad
n\ne n'\,.
\ee
By the angle brackets we denote the usual inner product in \l2$\!$
and the round brackets define a new inner product to be defined later
(see below).

{\bf II.}
The spectrum of $H$ coincides with the zeros of the determinant
$\Delta (E)$ given by (\ref{D})
or equivalently with the solutions of the equation $\psi(\pi,k)=0$
where $\psi(x,k)$ is a solution vanishing at $x=-\pi$ for all $k$.
If $H$ is Hermitian
$\Delta(E)$ has only simple zeros. For complex potentials the
equation $\Delta(E)=0$ may have multiple
zeros  as in the example above.
The situation with multiple zeros is an extension to differential
equations of the property known in the linear algebra for a
non-diagonalizable matrix which can nevertheless be reduced to a canonical Jordanian
form. Using this analogy one may call such Hamiltonians
{\em non-diagonalizable}. The set of their eigenfunctions is not
complete in $L^2(-\pi,\pi)$ but may be completed.
Otherwise the Hamiltonian is called
{\em diagonalizable} (cf. \cite{Wong}). For a diagonalizable
Hamiltonian $(\psi_n,\psi_n)\ne 0$ $\forall n$,
the set $\{\psi_n\}$ is complete in $L^2(-\pi,\pi)$ and they can always
be normalized such that
\be\label{norm}
(\psi_n,\psi_n)=\int_{-\pi}^\pi \psi_n^2(x)dx=1\,.
\ee
For the case when a continuous spectrum is present the concept of
diagonalizability should be examined more carefully.

{\bf III.}
In the conventional quantum mechanics the
property that self-adjoint operators have a complete set of
eigenfunctions in corresponding Hilbert space plays a crucial role. Now we would like
to discuss the role of this property in the case of  non-selfadjoint operators.
In particular, the following theorem (\cite{Naimark},
theorem 4, chapter II section 3) is extremely useful:
\begin{Th}
Let operator $H$ be generated by regular boundary conditions. Let
all its eigenvalues be simple zeros of the function $\Delta(E)$
defined in (\ref{D}).
Then any $f(x)$ belonging to the domain of definition of $H$ can be developed
over its eigenfunctions in the uniformly convergent series
\be
f(x)=\sum_{n=1}^\infty a_n\psi_n(x)
\ee
\be
a_n=\int_{-\pi}^{\pi}f(y)\wt\psi_n^*(y)dy
\ee
where $\psi_n(x)$, $\wt\psi_n(x)$ are eigenfunctions of $H$ and
$H^+$ corresponding to the eigenvalues $E_n$ and $E_n^*$
respectively.
  \end{Th}

  {\em Remark.} It is not explicitly stated in this
  theorem but the eigenfunctions are assumed to be
  normalized to satisfy Eq. (\ref{norm}).

We refer the reader to the book \cite{Naimark} for the general definition
of regular boundary conditions of a boundary value problem for an $n$th
order differential operator.
For our purposes it is sufficient to notice that
{\it nondegenerate}
boundary conditions used by Marchenko \cite{Marchenko} are
regular. They are specified as the conditions for which the
characteristic function $\Delta(E)$ is not constant for
the zero potential $V(x)=0$. Evidently the
 boundary conditions given in (\ref{bc}) satisfy this property.
 At first glance it would seem
that this is exactly the result that one needs in quantum mechanics. But for a
self-adjoint Hamiltonian a stronger theorem is valid, namely, the
set of its eigenfunctions is complete in \l2. This means that for
any element from \l2 corresponding Fourier series converges
in the squared mean.
A similar statement takes place for complex Hamiltonians also
but under some additional restriction imposed on the boundary
conditions. We will not go to further details but refer the
interested reader to the book by Naimark \cite{Naimark}.
Only we notice that
the non-degenerate boundary conditions by Marchenko and in
particular
the conditions (\ref{bc}) have this property and the
 eigenfunctions of the boundary value problem (\ref{1}-\ref{bc})
 form Riesz basis known also as a
basis equivalent to an orthonormal basis in which
case a counterpart of the
Parseval equality can be formulated for any element from \l2$\!$.
The completeness condition for the set of eigenfunctions of $H$
normalized according to (\ref{norm})
in the space \l2
has almost the usual form, only the
complex conjugation is absent
\be\label{cc}
\sum_{n=1}^\infty\psi_n(x)\psi_n(y)=\delta(x-y)\,.
\ee

Once the property that the system of eigenfunctions
of $H$ is complete in \l2 is established we can define
 a new Hilbert space as follows. First we define a new positive
 definite
 sesquilinear functional over the linear hull (lineal) ${\cal L}$
 of
all finite linear combinations of the
 solutions of the boundary value problem
 (\ref{1}-\ref{bc}) and then close this space with respect to the
 norm generated by this functional. For that we notice that
 together with the equation (\ref{1}) we have the adjoint
 boundary value problem defined by the differential equation
 $(H^+ - E)\wt\psi=0$ with the boundary conditions (\ref{bc}). In
 general, the lineal ${\cal L}^*$ of corresponding solutions of
 the latter
 equation is different from ${\cal L}$ although both they are
 included in \l2$\!$. According to the left hand side of Eq.
 (\ref{ip}) just elements from ${\cal L}^*$ participate in the
 biorthogonality condition. Therefore, to be able to use this
 equation while defining the new inner product
 for elements from ${\cal L}$
 we have to map
 lineal ${\cal L}^*$ onto ${\cal L}$. We realize this mapping
first between the basis functions,
$\psi_n\leftrightarrow \wt\psi_n=\psi_n^*$ and then continue it by linearity
to the whole spaces ${\cal L}^*$ and ${\cal L}$:
$\psi_n+\psi_m\leftrightarrow\wt\psi_n+\wt\psi_m=\psi_n^*+\psi_m^*$,
$c\psi_n\leftrightarrow c\wt\psi_n=c\psi_n^*$.
Once the
correspondence $\psi\leftrightarrow \wt\psi$,
is established $\forall\psi\in \cal L$,
$\forall\wt\psi\in {\cal L}^*$,
one can define in ${\cal L}$ a
positive definite sesquilinear
functional, $(\cdot ,\cdot )$,
(the new inner product) as follows:
\be\label{InPr}
(\psi,\vfi)=\langle \wt\psi|\,\vfi\rangle \qquad
\psi,\vfi\in{\cal L} \qquad \psi\rightarrow \wt\psi\in {\cal L}^*\,.
\ee
It is evident that because of the bi-orthonormality conditions
(\ref{ip},\ref{norm}) the basis $\psi_n$ becomes orthonormal with
respect to the new inner product.
Moreover, for any $\psi$ of the
form $\psi=\sum_{k=1}^nc_k\psi_k\ne 0$ the value
$(\psi,\psi)=\sum_{k=1}^n|c_k|^2$ being positive can be associated
with the squared norm, so that $\|\psi\|=(\psi,\psi)^{1/2}$. Now
the closure of the space $\cal L$ with respect to this norm gives
us the desired Hilbert space $\cal H$.
(The sesquilinear functional
 (\ref{InPr}) should be extended from the lineal ${\cal L}$ to the
 whole space ${\cal H}$ by continuity.)
 We point out that this construction is essentially based on the
 Hamiltonian $H$ and from this point of view the
 space  ${\cal H}$ is {\em dynamically defined} (cf. \cite{BBJ_PRD04}).

For multiple zeros of the function $\Delta(E)$ a complete
set in $L^2(-\pi,\pi)$
 can also be found but now together with the
eigenfunctions one has to find their associated functions.
For every eigenfunction $\psi(x,E_m)$ corresponding to a simple
eigenvalue $E_m$
with $p_m$ being the order of the zero $E_m$ of the function $\Delta(E)$
the chain of
associated functions is  defined as \cite{Naimark,Marchenko}:
$[\p^n\psi(x,E)/\p E^n]_{E=E_m}$, $n=1,\ldots ,p_m-1$.
The eigenfunctions together with the set of corresponding
associated functions form for the given value of the energy a root
subspace.
The completeness condition of
the set of eigenfunctions enlarged by
corresponding associated functions was first studied by
 Keldysh \cite{Keldysh,Naimark}.
 It is clear that in every $p_m$-dimensional root subspace one can
 choose a basis $\xi_{m_n}$ such that
 $(\xi_{m_n},\xi_{m_{n'}})=\delta_{m_n,m_{n'}}$ and the complete set of
 eigenfunctions and associated functions may be transformed into
 a set
 orthonormal with
 respect to the inner product  $(\cdot ,\cdot)$. Since our
 construction of the Hilbert space ${\cal H}$ is based on the
 orthonormality of the basis set $\{\psi_n\}$ with respect to this
 inner product  it is valid for the set $\{\xi_{m_n}\}$
 also.

{\bf 4.}
The properties of \pt-symmetric {\it diagonalizable} Hamiltonians
with unbroken \pt\ symmetry,
a real spectrum and the eigenfunctions normalized
according to (\ref{norm}), permit us to state that \\
{\bf A)}
the \cpt-inner product defined in \cite{BBJ}
coincides with the inner product $(\cdot ,\cdot )$ introduced
above.
This follows
from the fact that they apparently coincide on
solutions of the boundary value problem (\ref{1}-\ref{bc}) which
form a basis in \l2
and both are sesquilinear.
We infer, hence, that any such a Hamiltonian has
positive \cpt-normalizable eigenfunctions;
\\
{\bf B)}
\cpt\ completeness condition (see \cite{BBJ}) is equivalent to the usual
completeness condition for a non-selfadjoint
Hamiltonian in the space \l2$\!$ given by (\ref{cc});
\\
{\bf C)}
\cpt\ extension of quantum mechanics
for such Hamiltonians
should follow the same
lines already reported in \cite{BBJ} but as long as the situation with
the continuous spectrum is unclear this extension is incomplete.

We hope that
similar extension is possible for
Hamiltonians with continuous spectrum also.
This optimism is based on the fact
that for this case a
counterpart of the Parceval equality exists also
\cite{Naimark_DAN} but the lack of a counterpart of the
Riesz basis does not permit us to make a definite conclusion.

For non-diagonalizable Hamiltonians with a simple spectrum
 the situation is such that
for a given non-degenerate value of the energy there exist at least two
different functions, one is eigenfunction while the other is its
associated function. As our example shows in this case it is still
possible to construct a complete set orthonormal
with respect to an appropriately defined  inner product
 but the functions
belonging to the same root subspace are not eigenfunctions of the
Hamiltonian anymore.
Moreover, if the basis functions have a definite \pt\ parity the
\cpt\ inner product should coincide with the inner product
$(\cdot,\cdot)$ and, hence, the basis is positive
\cpt-normalizable.
We leave the question if the quantum
mechanics may be extended to this case open.

The work of BFS is partially supported by the
President Grant of Russia 1743.2003.2
 and the Spanish MCYT and European FEDER grant BFM2002-03773.
He would like to thank the staff of the Physics and
Applied Mathematics Unit of the Indian Statistical Institute for
hospitality in winter 2004 where this work has been started.
The authors would like to thank the referee for helpful comments.

\end{document}